\documentclass[twoside,onecolumn,12pt,longbibliography]{article}
\usepackage{extsizes}
\usepackage[super,sort&compress,comma]{natbib} 
\usepackage[version=3]{mhchem}
\usepackage[left=1.5cm, right=1.5cm, top=1.785cm, bottom=2.0cm]{geometry}
\usepackage[utf8]{inputenc}
\usepackage{amsmath}

\usepackage{physics}
\usepackage{makecell}
\usepackage{mathtools}
\usepackage{amsfonts}
\usepackage{amssymb}
\usepackage{graphicx}
\usepackage{wrapfig}
\usepackage{textcomp}
\usepackage{color}
\usepackage[dvipsnames]{xcolor}
\usepackage{multirow}
\usepackage{comment}
\usepackage{gensymb}
\usepackage{adjustbox}
\usepackage{chemformula}
\usepackage{longtable}
\usepackage{tabularx}
\usepackage{setspace}
\usepackage{sectsty}
\usepackage{droidsans}
\usepackage{charter}
\usepackage{titlesec}
\usepackage[format=plain,justification=justified,
singlelinecheck=false,font={stretch=1.125,small,sf},
labelfont=bf,labelsep=space]{caption}
\usepackage{balance}
\usepackage[colorlinks,bookmarks=true,citecolor=blue,linkcolor=red,urlcolor=blue]{hyperref}
\usepackage{orcidlink}

\sffamily
\title{Berry Curvature Dipole-induced Non-linear Hall Effect in Oxide Heterostructures}

\author{
  Nesta Benno Joseph,\textsuperscript{a,*} Arka Bandyopadhyay,\textsuperscript{b} Ajit C. Balram\orcidlink{0000-0002-8087-6015},\textsuperscript{c,d}\ Awadhesh Narayan\textsuperscript{a,\$} \\
  \\ \small
  \textsuperscript{a}Solid State and Structural Chemistry Unit, Indian Institute of Science, Bangalore 560012, India \\ \small
  \textsuperscript{b}Institute for Theoretical Physics and Astrophysics, University of W\"urzburg, D-97074 W\"urzburg, Germany
 \\ \small
 \textsuperscript{c}Institute of Mathematical Sciences, CIT Campus, Chennai 600113, India \\ 
 \small
 \textsuperscript{d}Homi Bhabha National Institute, Training School Complex, Anushaktinagar, Mumbai 400094, India\\
 \small \textsuperscript{*}nestajoseph@iisc.ac.in   
 \small  \textsuperscript{\$}awadhesh@iisc.ac.in
}

\date{}

\sectionfont{\sffamily\Large\mdseries}   
\subsectionfont{\normalsize\bfseries}    
\subsubsectionfont{\bfseries}            

\begin{document}

\maketitle

\begin{abstract}
The observation of non-linear Hall effects in time-reversal invariant systems has established the intriguing role of band topology beyond Berry curvature in determining transport phenomena. Many of these non-linear responses owe their origin to the Berry curvature dipole (BCD), which, like the Berry curvature (monopole), is also an electronic band structure effect, but is routinely strongly constrained by crystalline symmetries. Here, we propose non-centrosymmetric transition metal oxide heterostructures as promising platforms for realizing and tuning BCD-induced non-linear Hall effects. Specifically, we investigate superlattices of the form $(\mathrm{Ba(Os,Ir)}\mathrm{O}_3)_n/(\mathrm{BaTiO}_3)_4$ ($n{=}1, 2$), comprising metallic perovskite layers ($\mathrm{BaOsO_3}$ or $\mathrm{BaIrO_3}$) sandwiched between insulating ferroelectric $\mathrm{BaTiO_3}$ (BTO). The ferroelectric distortion in BTO breaks inversion symmetry of the superlattice, giving rise to a finite BCD with two symmetry-allowed components of equal magnitude and opposite sign. Our first-principles calculations demonstrate that the magnitude of the BCD -- and consequently the nonlinear Hall response -- can be effectively tuned by varying the number of metallic layers or the choice of the B-site cation in these \ch{ABO3} perovskites. Since Rashba splitting and ferroelectric distortion in these systems are readily controllable via an external electric field or strain, the non-linear Hall response in these materials can be directly engineered. Our findings establish non-centrosymmetric oxide perovskite heterostructures as a versatile platform for exploring and manipulating BCD-driven non-linear transport phenomena.
\end{abstract}

\doublespacing
\section*{\label{sec:oh_intro}Introduction}

Traditionally, Hall effects have been tied to the breaking of the time-reversal (TR) symmetry either due to an external magnetic field or the intrinsic magnetism of the material~\cite{Klitzing80, Haldane88, chien2013hall, cage2012quantum, chang2023colloquium, chang2016quantum}. However, the constraint that Hall conductance vanishes for TR invariant systems holds only in the linear response regime. Non-linear currents quadratic in the external field, referred to as non-linear Hall effect (NLHE)~\cite{du2021nonlinear,ortix2021nonlinear,bandyopadhyay2024non,suarez2025nonlinear}, can appear even in TR invariant systems. The key insight is that time-reversal symmetry breaking is not essential if the Hall response is considered in regimes beyond the linear order, provided the system has broken inversion symmetry~\cite{sodemann2015quantum}. This NLHE is determined by the Berry curvature dipole (BCD) tensor that measures the first moment of the Berry curvature [and not the Berry curvature (monopole) itself, as in the case of linear Hall effects] over the occupied states. This BCD gives rise to a second-order electrical Hall response induced by an electric field applied parallel to it.

The identification and study of materials that host BCD can provide a deeper understanding of its underlying nature, while offering fertile platforms to observe and engineer NLHEs to realize their potential applications, such as in terahertz sensors~\cite{zhang2021terahertz} and energy harvesting~\cite{kumar2021room, kumar2024quantum}. A rapidly growing list of materials has been identified, both theoretically and experimentally, as suitable candidates for exhibiting the NLHE. Experimental observations of BCD-induced NLHE includes, among others, the non-linear responses measured in van der Waals materials~\cite{ma2019observation, kang2019nonlinear, ma2022growth,huang2023giant, kang2023switchable, ho2021hall, huang2023intrinsic, he2022graphene, sinha2022berry, zhong2023effective}, topological crystalline insulators~\cite{zhang2022pst,nishijima2023ferroic}, organic compounds~\cite{kiswandhi2021observation} as well as Dirac and Weyl semimetals~\cite{shvetsov2019nonlinear, kumar2021room, min2023strong}. There have also been reports of NLHE measured at room temperature for a few systems~\cite{kumar2021room, min2023strong, makushko2024tunable, orlova2023gate, hu2023terahertz, lu2024nonlinear}.
Transition metal dichalcogenides~\cite{son2019strain, you2018berry, xiao2020two, zhou2020highly, joseph2021tunable, zhang2018electrically, he2021giant, jin2021strain}, topological semimetals~\cite{zhang2018berry, zeng2021nonlinear, singh2020engineering, pang2024tuning, chen2019strain}, graphene-analogs~\cite{bandyopadhyay2022electrically, bandyopadhyay2023berry}, and chiral systems~\cite{joseph2024chirality} are also among the materials that have been theoretically proposed to harbor BCD and exhibit NLHE. Aside from intrinsic BCD, extrinsic scattering effects can also result in NLHE under TR symmetric conditions in non-centrosymmetric materials~\cite{du2019disorder, du2021nonlinear, du2021quantum, ortix2021nonlinear, bandyopadhyay2024non}. 

Among the myriad of materials investigated, a thorough exploration of \emph{oxides} as potential candidates to observe NLHE is still lacking, although they constitute a large class of materials that may break inversion symmetry. Previous observations of NLH response in oxides were limited to two-dimensional electron gas (2DEG) confined at heterointerfaces, with the majority of the reports attributing the response to extrinsic contributions. 
The transverse NLH signals measured in the 2DEGs at the \ch{LaAlO3/KTaO3} (111) heterointerface~\cite{zhai2023large} and Si/SiO$_x$/ZnO interface~\cite{dashnonlinear}, and the light-induced enhancement of the NLH signal at (111)-oriented \ch{CaZrO3}/\ch{KTaO3}~\cite{zhang2025light} were found to be the result of electron skew-scattering, an extrinsic contribution to NLHE, rather than arising from the intrinsic BCD. 
In contrast, in \ch{LaAlO3/SrTiO3} interfaces grown along the [111] direction, very recent reports suggest that the distribution of orbital-sourced Berry curvature features both hotspots and singular pinch points, leading to a non-zero BCD~\cite{lesne2023designing, mercaldo2023orbital}, which should then result in an intrinsic NLHE.

\begin{figure*}[tbp]
    \centering
    \includegraphics[scale=0.12]{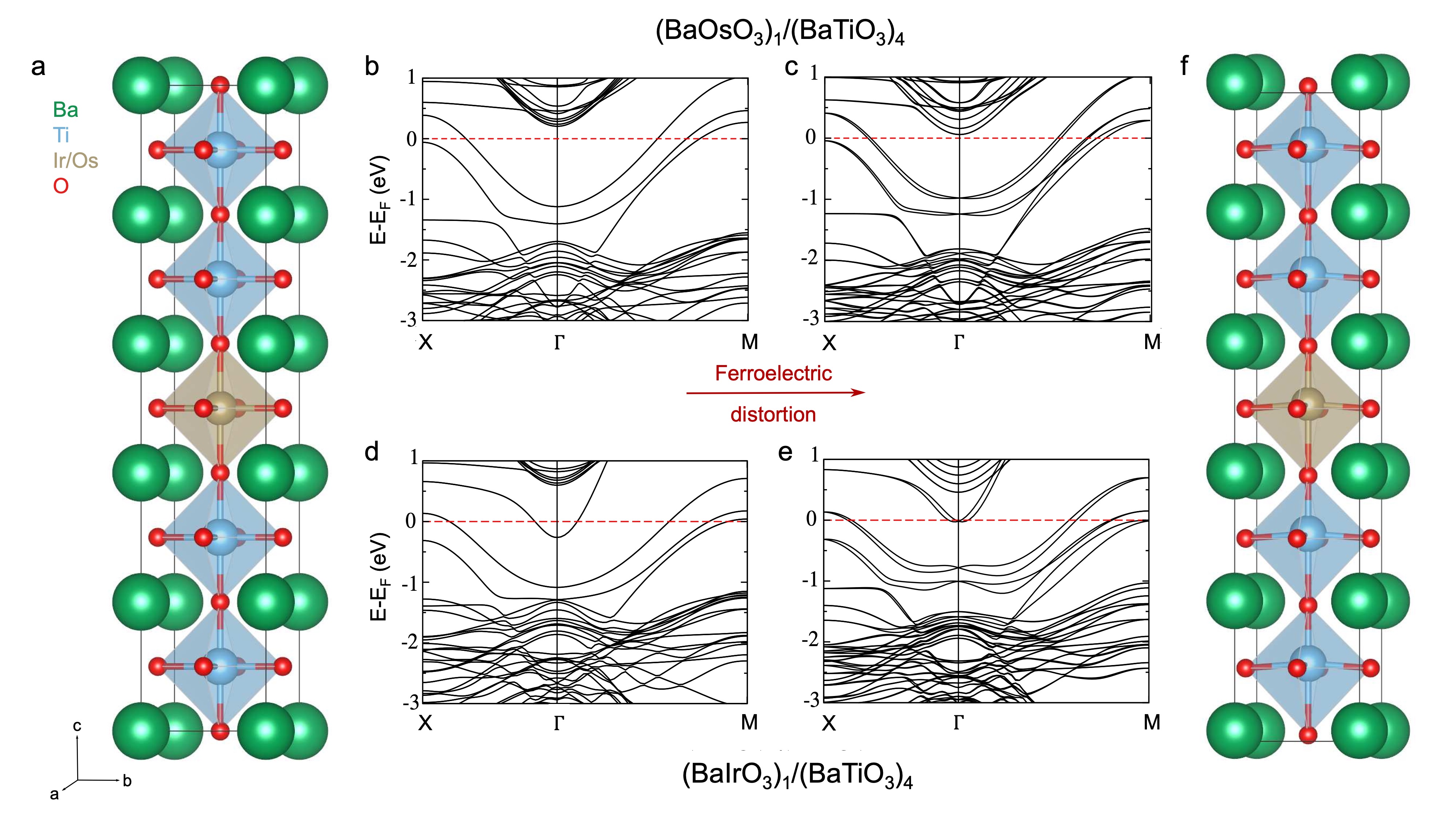}
    \caption{{\textbf{Centrosymmetric and non-centrosymmetric} $\mathbf{n{=}1}$\textbf{~heterostructures of \ch{(Ba(Os,Ir)O3)$_n$/(BTO)4}.}} The heterostructure (a) without and (f) with the ferroelectric distortion. The green, blue, brown, and red atoms represent Ba, Ti, Ir/Os, and O, respectively. 
    Electronic band structure of the centrosymmetric and non-centrosymmetric structures of (b)-(c) \ch{(BaOsO3)1/(BTO)4} and (d)-(e) \ch{(BaIrO3)1/(BTO)4}. The broken inversion symmetry leads to a large Rashba splitting of the bands. The bands around the Fermi level (red dashed line) arise predominantly from the $d$ orbitals of Os/Ir.}
    \label{fig:1l_bands}
\end{figure*}

In this work, we propose transition metal oxide heterostructures as promising candidates for realizing intrinsic BCD-induced NLHE. Such heterostructures are different from the previously examined oxide heterointerfaces -- these are superlattices of the form \ch{(ABO3)$_n$/(AB'O3)$_m$} having the same A-site cation and two different B-site cations sharing the O octahedra.
The candidate materials are constructed by building a multilayer of metallic perovskites, either \ch{BaOsO3} or \ch{BaIrO3}, sandwiched between the well-known ferroelectric insulator \ch{BaTiO3} (BTO), leading to the formation of \ch{(Ba(Os, Ir)O3)$_n$}/(BTO)$_4$ heterostructures, where $n{=}1, 2$, denote the number of metallic perovskite layers [see Fig.~\ref{fig:1l_bands}(a)]. We analyze the symmetries associated with these heterostructures to find that only two components of the BCD tensor are symmetry-allowed to be non-zero; moreover, these have to be equal in magnitude and opposite in sign in these superlattices. Our first-principles calculations of the Berry curvature distribution and the corresponding BCD reveal that both the number of metallic layers and the choice of B-site cation can effectively tune the NLH response in these systems. Our work establishes oxide perovskite heterostructures as viable candidates to explore intrinsic BCD-induced NLHE.


\section*{\label{sec:oh_methods} Computational Details}

Our first-principles calculations were carried out using the DFT framework as implemented in the {\sc quantum espresso} code~\cite{QE-2017, QE-2009}. We employed the Perdew, Burke, and Ernzerhof (PBE)~\cite{perdew1996generalized} form of generalized gradient approximation (GGA) for the exchange-correlation functional, and ultra-soft pseudopotentials~\cite{PhysRevB.41.7892} to describe the core electrons.

For this study, four different heterostructures were considered -- named henceforth -- the $n{=}1$ heterostructures \ch{(BaOsO3)1/(BTO)4} and \ch{(BaIrO3)1/(BTO)4}, and the $n{=}2$ heterostructures \ch{(BaOsO3)2/(BTO)4} and \ch{(BaIrO3)2/(BTO)4}, depending on the number of layers of the metallic perovskite. The heterostructures were constructed from a $5{\times}1{\times}1$ supercell of BTO, replacing one of the five Ti atoms with either Os or Ir and allowing the system to relax until all forces were less than at least $10^{-4}$ eV/\AA.
A kinetic energy cut-off of 65 Ry was considered for all four systems, with the Brillouin zone sampled over a uniform $\Gamma$-centered $k$-mesh of $8{\times}8{\times}4$, including spin-orbit coupling (SOC).

Wannier-based tight-binding (TB) model computations were carried out by constructing maximally localized Wannier functions (MLWFs) using the {\sc wannier90} code~\cite{mostofi2014updated}, with O $p$, and Os/Ir and Ti $d$ orbitals as the basis. 
From the constructed MLWFs, the BCD can be numerically evaluated as detailed below. 

\noindent
The BCD component, $D_{ab}$, is~\cite{sodemann2015quantum}
\begin{equation}
    D_{ab} = \sum_{n}\int_{k}~f_n^0(\textbf{k})~\frac{\partial\Omega^{(n)}_b}{\partial k_a}d\textbf{k},
    \label{eqn:bcdexpression}
\end{equation}
where $f_n^0$(\textbf{k}) is the equilibrium Fermi-Dirac distribution and $\Omega$ is the Berry curvature, given by~\cite{Thouless82}
\begin{equation}
    \Omega^{(n)}_a(\textbf{k}){=}\varepsilon_{abc} \partial_{b}\mathcal{A}^{(n)}_{c}(\textbf{k}).
    \label{eqn:berry_curv}
 \end{equation}   
Here, $\varepsilon_{abc}$ is the Levi-Civita tensor, which is completely antisymmetric under exchange of any pair of indices and takes values ${+}1$, ${-}1$, or $0$ depending on whether $a,b,c \; {\in} \;\{x,y,z\}$ is an even permutation, odd permutation, or contains repeated indices, respectively. The quantity $\partial_b{\equiv}\frac{\partial}{\partial k_b}$ denotes the partial derivative with respect to the $b$-th component of the crystal momentum vector $\textbf{k}$. The Berry connection $\mathcal{A}^{(n)}_{c}(\textbf{k})$ is given by
\begin{equation}    
    \mathcal{A}^{(n)}_{c}(\textbf{k}){=}\langle u_{n}(\textbf{k}) |i\grad_{\textbf{k}}| u_{n}(\textbf{k}) \rangle,
\end{equation}
where $|u_{n}(\textbf{k})\rangle$ is the periodic part of the Bloch wavefunction corresponding to the $n$-th band, and $\nabla_{\textbf{k}}$ denotes the gradient with respect to the momentum vector $\textbf{k}$. The Berry connection can be interpreted as a gauge potential in momentum space, analogous to the vector potential in electromagnetism, while the Berry curvature plays the role of an effective magnetic field in $\textbf{k}$-space.

The crystal point symmetries of the materials impose additional constraints on the BCD tensor $D$, which, as per Neumann's principle, take the form~\cite{sodemann2015quantum}
\begin{equation}\label{eqn:symmetry_const}
    D = \mathrm{det}(S)SDS^T,
\end{equation}
where $S$ is the orthogonal matrix that describes the point symmetry and $\mathrm{det}(S)$ denotes the determinant of the matrix $S$. The non-zero components of BCD can be identified by analyzing the condition stated in Eq.~\eqref{eqn:symmetry_const}. We used our Wannier-based TB Hamiltonian to compute the Berry curvature and associated BCD using the {\sc wannier-berri} code~\cite{tsirkin2021high}. Our Berry curvature and BCD calculations were done at a temperature of 40 K, on $k$-grid sizes of $26{\times}26{\times}4$ and $126{\times}126{\times}21$, respectively. Note that increasing this temperature broadens the features in the BCD as a function of the Fermi level.


\section*{\label{sec:oh_results}Results and Discussion}

\subsection*{Structure and Symmetry}

The material candidates under study are constructed such that one or two layers of the metallic perovskite, either \ch{BaOsO3} or \ch{BaIrO3}, alternate between four layers of BTO, forming a heterostructure of the form shown in Fig.~\ref{fig:1l_bands}(a) for $n{=}1$.  
BTO is a well-known and well-studied ferroelectric material, while the metallic \ch{BaOsO3} synthesized under high pressure has a cubic crystal structure~\cite{shi2013high}. Among the different phases of BTO, the high-temperature ($T{\approx}400$ K) paraelectric phase has the ideal perovskite structure, with the Ba atoms occupying the corners and the Ti atoms occupying the center of the cubic lattice, with the latter surrounded by an oxygen octahedron. This phase of the heterostructure will not host BCD, since it is inversion symmetric.
The ferroelectric phase of BTO at room temperature has an out-of-plane distortion of the Ti atoms, leading to a broken inversion symmetry in the system. On the construction of heterostructures, as described previously, starting from the paraelectric structure of BTO, the system remains inversion symmetric [Fig.~\ref{fig:1l_bands}(a)]. In contrast, the heterostructure, with BTO in the ferroelectric phase, is non-centrosymmetric since the out-of-plane distortion of Ti atoms in the BTO layer \textit{induces} a distortion of the central Os/Ir atoms because of the shared O octahedra [Fig.~\ref{fig:1l_bands}(f)]. 

This is evident from the electronic structure of these systems as well.
The construction of such heterostructures leads to the thin metallic layer of \ch{Ba(Os, Ir)O3} trapped between the insulating BTO~\cite{zhong2015giant}. 
As BTO is a ferroelectric insulator with a bandgap of $\sim3.2$~eV, the metallicity in the system is exclusively due to the presence of the metallic perovskite layer.
Fig.~\ref{fig:1l_bands}(b)-(e) show the band structures of (\ch{BaOsO3)}$_1$/(BTO)$_4$ and (\ch{BaIrO3)}$_1$/(BTO)$_4$ heterostructures, where the three pairs of bands around Fermi level have their origin from the B-site cation (Os/Ir) $d$ orbitals of the metallic layer, specifically the $t_{2g}$ orbitals $d_{xy}, d_{yz}$ and $d_{xz}$. 
The broken inversion symmetry in the ferroelectric distorted heterostructures coupled with SOC interactions leads to a giant Rashba splitting in both \ch{(BaOsO3)1/(BTO)4} and \ch{(BaIrO3)1/(BTO)4} as shown in Fig.~\ref{fig:1l_bands}(c) and (e). This is in contrast to the band structures of the inversion symmetric systems [Fig.~\ref{fig:1l_bands}(b) and (d)], where no such splitting is discernible due to inversion symmetry being maintained.
As we discussed earlier, broken inversion symmetry is the necessary condition to observe BCD-induced NLHE, and hence, our proposed non-centrosymmetric systems are ideal for the investigation of BCD from a symmetry perspective. 
All our discussions henceforth are focused on these distorted, inversion-broken heterostructures.

The number of non-zero components of BCD will depend on the other symmetries present in the system. The distorted heterostructure has a space group of $P4mm$ (as opposed to $P4/mmm$ of the centrosymmetric structure), with C$_2$ and C$_4$ rotations about the $z$ axis, and four mirror planes -- $M_x, M_y, M[110]$ and $M[1\bar{1}0]$. Imposing these symmetry constraints on the BCD tensor of the non-centrosymmetric system, as mandated by Eq.~\eqref{eqn:symmetry_const}, allows for only two of the BCD components to be non-zero and also imposes a relation between them. Thus, the BCD tensor is
\begin{equation}
\label{eqn:bcd_tensor}
D =
\begin{bmatrix}
0       & D_{xy} & 0 \\
D_{yx} & 0      & 0 \\
0       & 0      & 0
\end{bmatrix},
\end{equation}
with the relation $D_{yx}{=}{-}D_{xy}$.

Hence, according to Eq.~\eqref{eqn:bcdexpression}, the $\Omega_x$ and $\Omega_y$ components of the Berry curvature lead to the presence of BCD in these systems, as we shall discuss next in the following sections. 

\subsection*{\ch{(Ba(Os,Ir)O3)1/(BaTiO3)4} heterostructures}

\begin{figure*}[!tbp]
    \centering
    \includegraphics[scale=0.125]{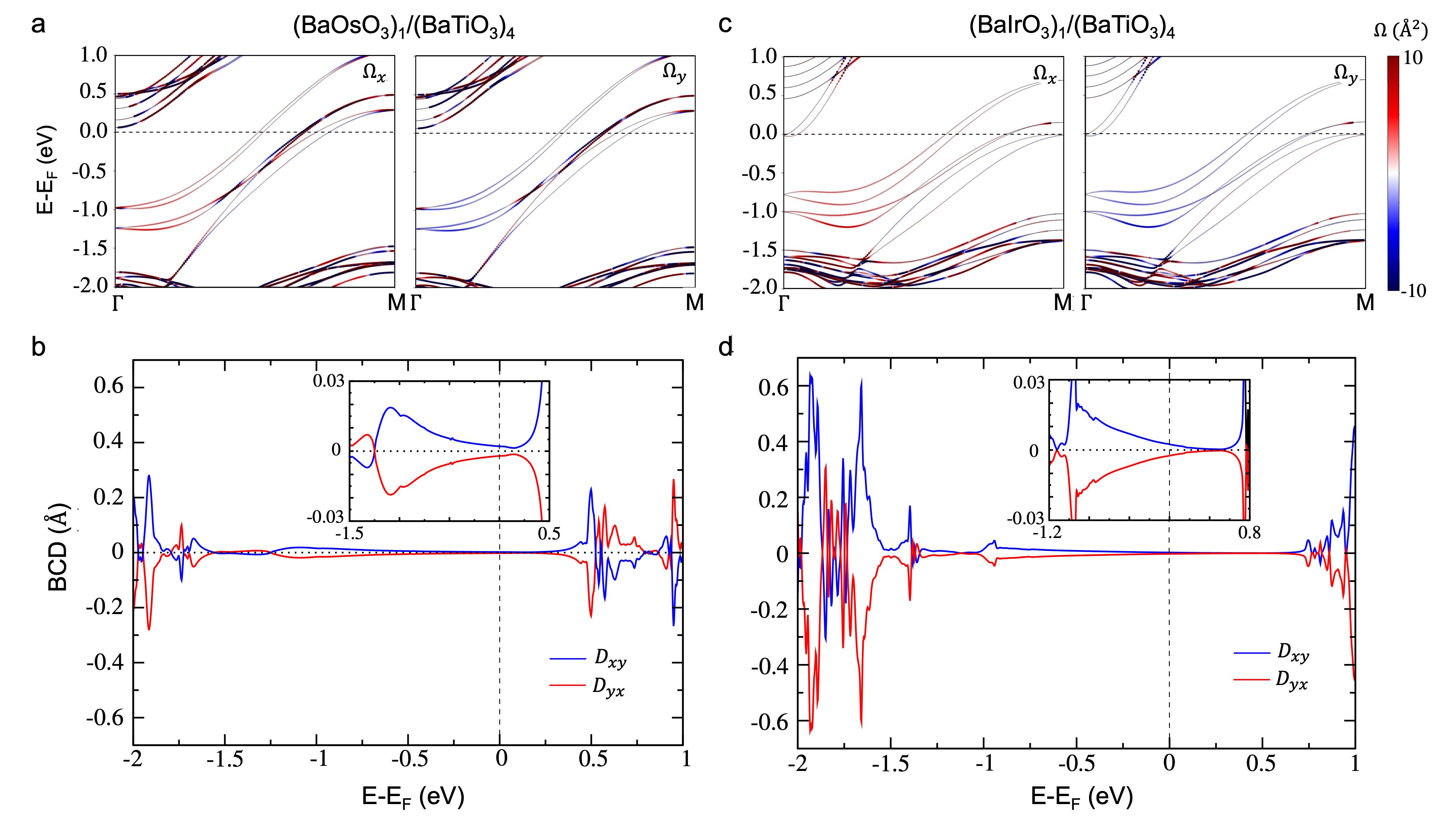}
    \caption{\textbf{Berry curvature and BCD of $\mathbf{n{=}1}$ heterostructures.} Berry curvature components $\Omega_x$ and $\Omega_y$, superimposed on the band structure along high-symmetry $\Gamma-M$ direction, associated with (a) \ch{(BaOsO3)1/(BTO)4} and (c) \ch{(BaIrO3)1/(BTO)4}. The red and blue colors represent positive and negative values of the Berry curvature components, respectively. The two equal and opposite components, $D_{xy}$ and $D_{yx}$, of the BCD as a function of the energy with respect to the Fermi level associated with (b) \ch{(BaOsO3)1/(BTO)4} and (d) \ch{(BaIrO3)1/(BTO)4}. The inset shows the components magnified around $E{-}E_F{=}0$.} 
    \label{fig:1l_bcd}
\end{figure*}

The first set of candidate materials that we consider is the $n{=}1$ heterostructures, where one layer of either \ch{BaOsO3} or \ch{BaIrO3} is alternating between four layers of BTO. 
Focusing first on \ch{(BaOsO3)1/(BTO)4} system, Fig.~\ref{fig:1l_bcd}(a) shows the $x$ and $y$ components of the Berry curvature, $\Omega$, superimposed onto the band structure of the system, along the high symmetry $\Gamma - M$ path.
As mentioned above, the metallicity in these heterostructures arises from the $d$ orbitals of Os. 
These metallic bands are less dispersive in the $z$ direction, compared to other directions, because along the $z$-axis, the metallic layer is confined between the insulating layers. 
The Berry curvature distribution along the bands around the Fermi level is mostly uniform, with local concentrations and rapid variations near regions of band anticrossings. 
Upon computing the non-zero $D_{xy}$ and $D_{yx}$ components of BCD, we find that the two components are equal and opposite to each other, consistent with the symmetry analysis resulting in Eq.~\eqref{eqn:bcd_tensor}. 
The low density of states around the Fermi level and uniformity in the Berry curvature distribution are directly reflected in the BCD, plotted as a function of chemical potential in Fig.~\ref{fig:1l_bcd}(b).
While the values of BCD obtained are not very high compared to some of the other predicted materials, nevertheless, we find non-zero BCD associated with the \ch{(BaOsO3)1/(BTO)4} heterostructure around the Fermi level.
We will subsequently examine potential approaches to enhance these values.

Replacing the Os with Ir, we find that the Rashba splitting is larger owing to the larger SOC effect in Ir, as seen in Fig.~\ref{fig:1l_bands}(e) and Fig.~\ref{fig:1l_bcd}(c). 
As observed in the case of the first heterostructure, the uniform distribution of Berry curvature associated with the Ir $d$ bands around the Fermi level is directly reflected in the BCD associated with the material [Fig.~\ref{fig:1l_bcd}(d)].
The non-zero BCD for both the $n{=}1$ heterostructures indicates that such systems can indeed display BCD-induced NLH response.

\subsection*{\ch{(Ba(Os,Ir)O3)2/(BaTiO3)4} heterostructures}

\begin{figure*}[!tbp]
    \centering
    \includegraphics[scale=0.125]{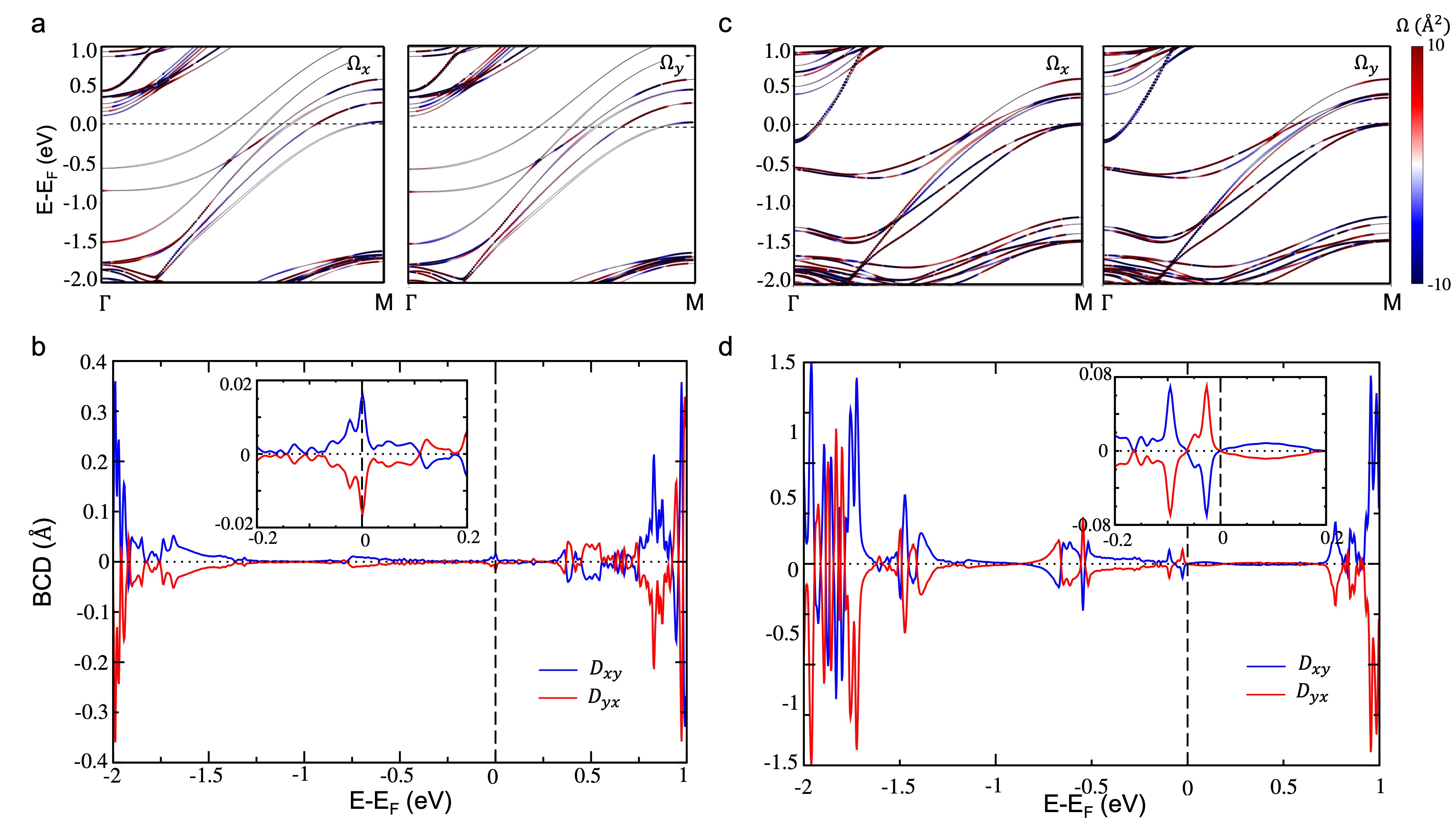}
    \caption{\textbf{Berry curvature and BCD of $\mathbf{n{=}2}$ heterostructures.} The Berry curvature components $\Omega_x$ and  $\Omega_y$, superimposed on the band structures along $\Gamma-M$ direction of (a) \ch{(BaOsO3)2/(BTO)4} and (c) \ch{(BaIrO3)2/(BTO)4}. The red and blue colors represent positive and negative values of the Berry curvature components, respectively. (b), (d) The non-zero BCD components $D_{xy}$ and $D_{yx}$ associated with the $n{=}2$ heterostructures, plotted as a function of energy with respect to the Fermi level. The inset shows a magnified region of BCD around $E-E_F=0$. Sandwiching two layers of \ch{(Ba(Os, Ir)O3)} between the BTO layers significantly enhances the BCD around the Fermi level. } 
    \label{fig:2l_bcd}
\end{figure*}

Since the set of bands around the Fermi level is exclusively due to the Os/Ir ions, one promising method to tune the NLH response is to increase the number of metallic layers. 
In this regard, we consider the second set of candidate systems -- $n{=}2$ heterostructures -- with two layers of metallic perovskite alternating between four layers of BTO. 
This increases the density of states within the insulating gap of BTO, arising from the set of $t_{2g}$ orbitals of Os/Ir. 
The band structure of the heterostructures along the $\Gamma-M$ direction, with the Berry curvature components superimposed on it, is presented in Fig.~\ref{fig:2l_bcd}(a) and (c). 
Only the $\Omega_x$ and $\Omega_y$ components are shown since, as discussed previously, only the $D_{xy}$ and $D_{yx}$ components of the BCD are non-zero. 

The increased number of band crossings leads to a larger number of Berry curvature hotspots in the $n{=}2$ heterostructures as compared to the $n{=}1$ case. This is also directly reflected in the BCD distribution as seen in Fig.~\ref{fig:2l_bcd}(b) and (d) of the two systems. 
Remarkably, we find that the BCD associated with \ch{(BaOsO3)2/(BTO)4} at the Fermi level is \emph{an order of magnitude larger} than its $n{=}1$ analog. We also find that, in the case of \ch{(BaIrO3)2/(BTO)4}, the BCD components exhibit a peak and a dip very close to the Fermi level, with a similar enhancement in magnitude as in the case of \ch{(BaOsO3)2/(BTO)4}. The fluctuations in the BCD can further be directly related to the distribution of the Berry curvature along the bands.

\begin{figure*}[!tbp]
    \centering
    \includegraphics[width=0.85\linewidth]{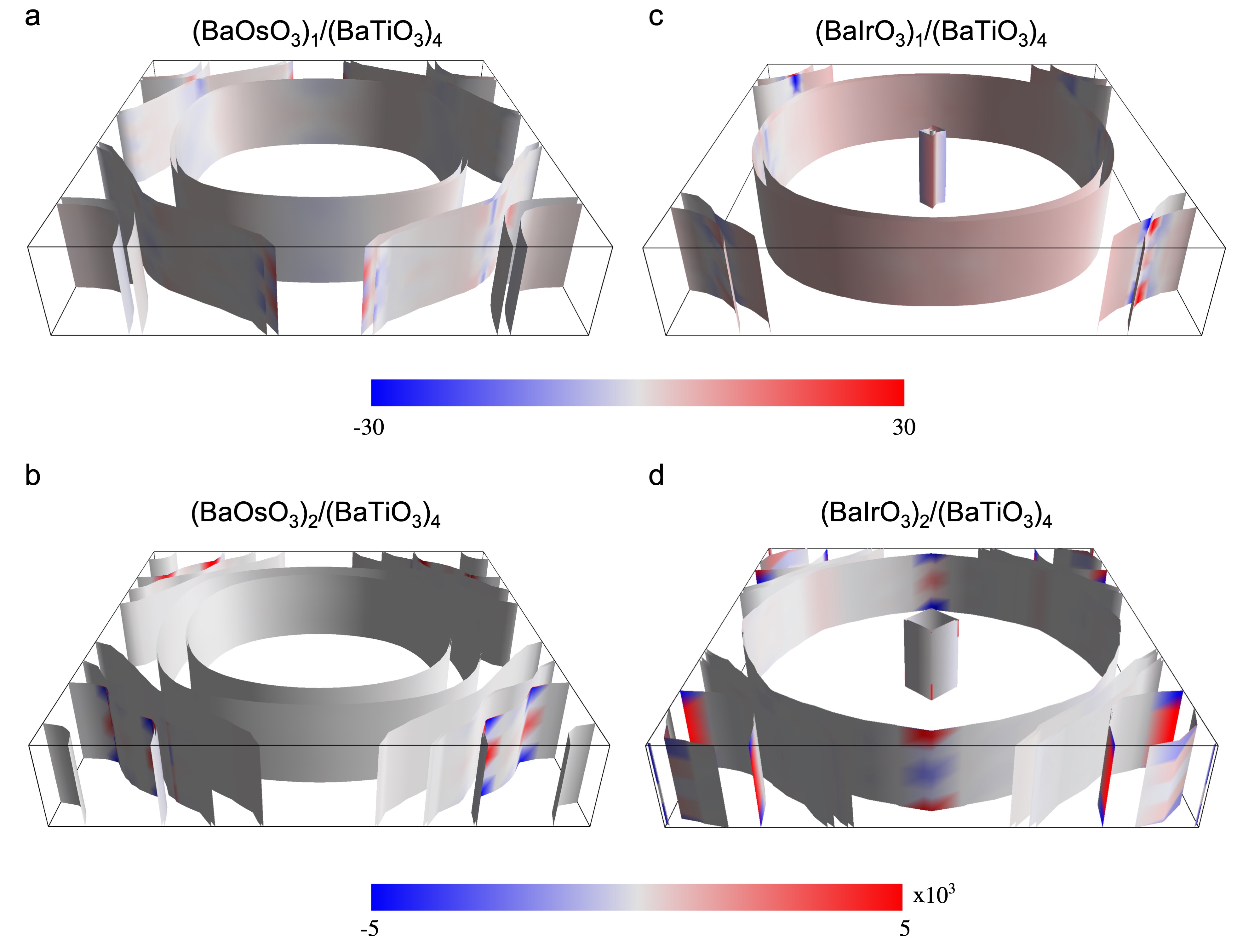}
    \caption{\textbf{A comparison of the distribution of the derivative of the Berry curvature.} The distribution of the quantity $\partial_x\Omega_y$ on the Fermi surface of (a) \ch{(BaOsO3)1/(BTO)4}, (b) \ch{(BaIrO3)1/(BTO)4}, (c) \ch{(BaOsO3)2/(BTO)4}, and (d) \ch{(BaIrO3)2/(BTO)4}. The red and blue colors indicate the positive and negative values, respectively. Note that the value of $\partial_x\Omega_y$ for $n{=}2$ heterostructures is two orders of magnitude higher than the $n{=}1$ case.
    }
    \label{fig:derberry}
\end{figure*}

To further understand the origin of the BCD, in Fig.~\ref{fig:derberry}, we present the distribution of the derivative of the Berry curvature on the Fermi surface for the four heterostructures that we proposed. 
This quantity -- the derivative of the Berry curvature -- weighted by the Fermi occupation function gives the BCD density, which, when integrated over the Brillouin zone, is the BCD [see Eq.~\eqref{eqn:bcdexpression}].
In Fig.~\ref{fig:derberry} we plot $\partial_x\Omega_y$, which gives rise to the $D_{xy}$ component of the BCD. Since $D_{xy}{=}{-}D_{yx}$, $\partial_y\Omega_x$ will have exactly the opposite distribution of $\partial_x\Omega_y$.
Examining the distribution of the derivative of the Berry curvature yields a number of insights.
First of all, the Fermi surfaces for the heterostructures based on Os or Ir are quite similar except for the small electron pockets centered at $\Gamma$.
These are absent for the Os based superlattices.
Furthermore, the larger central electron pocket around $\Gamma$ does not show any significant warping effects. As we increase the number of metallic layers, the number of Fermi sheets is significantly enhanced, from 6 in the $n{=}1$ heterostructures, to 12 and 16 for the $n{=}2$ heterostructures \ch{(BaOsO3)2/(BTO)4}, and \ch{(BaIrO3)2/(BTO)4}, respectively. This is expected due to the enhanced density of states within the insulating gap of BTO.
Comparing between Os and Ir-based superlattices, we find that the derivative of the Berry curvature is consistently larger for the Ir-based systems.
This behavior can be attributed to the stronger SOC-induced splitting in the presence of Ir relative to Os, as evident from Fig.~\ref{fig:1l_bands}, where a more pronounced Rashba splitting is observed in the $d$ orbitals of Ir. Notably, we find that the value of $\partial_x\Omega_y$ for $n{=}2$ heterostructures is two orders of magnitude larger than the $n{=}1$ cases.
This, in conjunction with an enhanced density of states, is directly reflected in the enhanced BCD in the $n{=}2$ heterostructures compared to the $n{=}1$ ones. Since the Berry curvature is also predominantly concentrated around such regions, the quantity $\partial_x\Omega_y$ is also mostly concentrated around the regions of band anticrossings.

Our observations on the distribution of the derivative of the Berry curvature over the Fermi surface allow us to suggest potential strategies to further engineer the BCD and the resulting NLH response.
In general, an enhanced BCD is expected with a higher density of states and an increased number of band anticrossings, which lead to Berry curvature hot-spots.
Furthermore, larger SOC appears to be helpful in engineering such band anticrossings. 
Given the variety of oxide materials available and the flexibility in creating their heterostructures, our predictions can be helpful in the search for promising platforms for tunable NLH response.
We note that the sodium analog of \ch{BaIrO3}, namely sodium iridate, \ch{Na2IrO3}, has a sizable intrinsic SOC~\cite{shitade2009quantum, Rod15_spin_textures_helical_edge_states}, making it another promising candidate to observe NLHEs. Here, we have focused on the Ba variant for its ease of heterostructure engineering with \ch{BaTiO3}, as \ch{Na2IrO3} forms a honeycomb structure.

\section*{\label{sec:oh_sum}Conclusions}

In this study, we proposed oxide heterostructures, sandwiching thin layers of metallic perovskites between the ferroelectric insulator BTO, as possible candidates for realizing BCD-induced NLHE. We considered four systems -- $n{=}1$ and $n{=}2$ heterostructures of \ch{(Ba(Os, Ir)O3)$_n$/(BTO)4} -- depending on the number of the metallic perovskite layers and the type of B-site metal cation.
We found that in such heterostructures, the ferroelectric distortion present in BTO induces a distortion of the B-site cation in the metallic layer, leading to broken inversion symmetry, which results in a BCD and, consequently, an intrinsic NLH response. Using first-principles calculations, we found that the symmetries present in such heterostructures only allow two components of the BCD to be non-zero -- $D_{xy}$ and $D_{yx}$ -- which are equal and opposite to each other. Encouragingly, increasing the number of metallic layers from one to two drastically changes the nature of the BCD around the Fermi level, leading to a BCD-induced NLH effect that is tunable with respect to the number and type of metallic B-site cation in such heterostructures. We note that the Rashba splitting in such heterostructures can be tuned by an electric field as well as strain~\cite{zhong2015giant}. The ferroelectric distortion of BTO may also be controlled using an electric field. Combining these different effects allows for a wide control and tunability of the NLH response in such systems. Other interesting non-linear effects, such as a current-induced gap opening that persists down to zero temperature, could also be observable in these systems~\cite{Balram19CurrentInducedGapOpeningTIs}. Thus, our predictions put forth non-centrosymmetric oxide heterostructures as suitable candidates to realize non-linear responses.

\section*{Acknowledgements}
We thank B. Jalan, M. Jain, H. R. Krishnamurthy, D. Varghese, E. Simmen, S. Bhowal, and N. Spaldin for valuable discussions and related collaborations. N.B.J. acknowledges the Prime Minister’s Research Fellowship (PMRF) for support. A.N. acknowledges support from DST CRG grant (CRG/2023/000114). 

\bibliography{references.bib}
\bibliographystyle{rsc}

\end{document}